# EXTRASOLAR PLANET TRANSITS OBSERVED AT KITT PEAK NATIONAL OBSERVATORY

BY

PEDRO V. SADA [1, 8], DRAKE DEMING [2, 8], DONALD E. JENNINGS [3, 8], BRIAN K. JACKSON [3, 8], CATRINA M. HAMILTON [4, 8], JONATHAN FRAINE [2, 8], STEVEN W. PETERSON [5], FLYNN HAASE [5], KEVIN BAYS [5], ALLEN LUNSFORD [3, 6, 8], EAMON O'GORMAN [7]

[1] Universidad de Monterrey, Departamento de Física y Matemáticas, Av. I. Morones Prieto 4500 Pte., San Pedro Garza García, Nuevo León, 66238, México
pedro.valdes@udem.edu.mx

[2] Department of Astronomy, University of Maryland, College Park, MD 20742, USA
ddeming@astro.umd.edu, jfraine@astro.umd.edu

[3] Planetary Systems Laboratory, Goddard Space Flight Center, Mail Code 693, Greenbelt, MD 20771, USA
donald.e.jennings@nasa.gov, brian.k.jackson@nasa.gov

[4] Dickinson College, Carlisle, PA 17013, USA
hamiltoc@dickinson.edu

[5] Kitt Peak National Observatory, National Optical Astronomy Observatory, 950 N Cherry Ave., Tucson, AZ, 85719, USA
speterson@noao.edu, fhaase@noao.edu, kbays@noao.edu

[6] The Catholic University of America, Washington, DC 20064, USA
allen.w.lunsford@nasa.gov

[7] Trinity College, Dublin, Dublin 2, IRELAND
eogorma@tdc.ie

[8] Visiting Astronomer, Kitt Peak National Observatory, National Optical Astronomy Observatory, which is operated by the Association of Universities for Research in Astronomy under cooperative agreement with the National Science Foundation.

# ABSTRACT

We obtained J-, H- and JH-band photometry of known extrasolar planet transiting systems at the 2.1-m Kitt Peak National Observatory Telescope using the FLAMINGOS infrared camera between October 2008 and October 2011. From the derived lightcurves we have extracted the mid-transit times, transit depths and transit durations for these events. The precise mid-transit times obtained help improve the orbital periods and also constrain transit-time variations of the systems. For most cases the published system parameters successfully accounted for our observed lightcurves, but in some instances we derive improved planetary radii and orbital periods. We complemented our 2.1-m infrared observations using CCD z'-band and B-band photometry (plus two Hydrogen Alpha filter observations) obtained with the Kitt Peak Visitor's Center telescope, and with four H-band transits observed in October 2007 with the NSO's 1.6-m McMath-Pierce Solar Telescope.

The principal highlights of our results are: 1) our ensemble of J-band planetary radii agree with optical radii, with the best-fit relation being: $(R_p/R_*)_J = 0.0017 + 0.979 (R_p/R_*)_{optical}$, 2) We observe star spot crossings during the transit of WASP-11/HAT-P-10, 3) we detect star spot crossings by HAT-P-11b (Kepler-3b), thus confirming that the magnetic evolution of the stellar active regions can be monitored even after the Kepler mission has ended, and 4) we confirm a grazing transit for HAT-P-27/WASP-40. In total we present 57 individual transits of 32 known exoplanet systems.

1. INTRODUCTION

Many exoplanet systems contain Jupiter-mass planets on close-in orbits. These planets are strongly irradiated by their host stars, and emit significant radiation in the infrared (Charbonneau et al. 2005, Deming et al. 2005). Characterization of their atmospheres using transit and secondary eclipse techniques has become a very active field (Seager and Deming 2010). Atmospheric observations using secondary eclipse are also sensitive to the orbital dynamics, specifically the eccentricity of the orbit, via the phase of the eclipse (Deming et al. 2007). Consequently, interpreting secondary eclipse observations requires knowing the ephemeris of the transits to high precision. Continuing explorations of discoveries by transit surveys have given us a sample of more than 70 hot Jupiters transiting systems brighter than V=13, and the increasing sample size makes it difficult to maintain accurate parameters for all systems. Some systems already require additional transit observations in order to attain sufficiently precise ephemeredes to interpret secondary eclipse phase (e. g., Todorov et al. 2011). Moreover, the discovery photometry for transiting planets typically provides only relatively coarse photometric precision, and follow-up photometry with larger aperture telescopes is needed to determine the giant planet radius to a precision limited only by knowledge of the stellar mass (e. g., Winn et al. 2007a). For these reasons, several groups are monitoring known transiting planets using moderate to large aperture telescopes (e. g., Southworth et al. 2009).

In addition to the motivations discussed above, we are interested in transit monitoring at near-IR (JHK) wavelengths. Near-IR wavelengths offer reduced stellar limb-darkening, and thereby provide an independent check on planetary radii inferred using optical photometry. Moreover, the composition of exoplanetary atmospheres may cause real differences in transit radii with wavelength that could eventually be detected with sufficiently precise observations. The relatively large sample of known transiting systems means that a general transit survey is well matched to classical observing and telescope scheduling methods. In this paper, we present results from a series of classical observing runs, producing high-precision photometry of several known exoplanet transiting systems observed in the near infrared (J-, H- and JH-bands), and several less precise transits in the optical. Our photometry was obtained at the Kitt Peak National Observatory using several telescopes. In total, we present 57 lightcurves of 32 transiting exoplanetary systems. In section 2 we present the observations and data analysis methods used, while in section 3 we describe fitting to the photometry using Markov Chain Monte Carlo (MCMC) methodology (Ford, 2005). Section 4 discusses results derived from the model fits, by comparing our results for planetary radii with results in the optical, and we discuss details of individual systems such as improvements to the orbital periods.

## 2. OBSERVATIONS AND PHOTOMETRY

### 2.1 OBSERVATIONS

Our primary observational system is the Kitt Peak National Observatory 2.1-meter reflector with the FLAMINGOS 2048×2048-pixel infrared imager and a J-band (1.25 μm), an H-band (2.50 μm), and a J- and H-band combination (JH) filters (Elston 1998). The 0.6 arc-sec per pixel scale yielded a FOV of ~20×20 arc-min with sufficient comparison stars available for differential photometry. Following the conclusion of nightly public programs, we also had access to the 0.5-meter telescope at the Kitt Peak Visitor Center (the VC telescope). With this telescope we used a 3072x2048 CCD camera at 0.45 arc-sec per pixel and typically a z'-band filter (although in a few instances the data was acquired through a B filter, and twice through a Hydrogen Alpha filter). Most of the observations presented here were obtained between October 2008 and October 2011. We also present four other transits obtained in October 2007 using the National Solar Observatories' (NSO) 1.6-meter McMath-Pierce Solar Telescope and their NSO Array Camera (NAC). The NAC is a cryogenically cooled 1K×1K InSb Aladdin-III array (Ayres et al. 2008). For these observations only one quadrant of the chip was used with only one comparison star available due to the small FOV (~5×5 arc-min).

Observations at all three telescopes used various degrees of defocus to improve the photometric precision, and all used automatic off-axis or manual guiding to maintain pointing stability: Exposure times varied between 20 seconds and 120 seconds, depending on the system used and the stellar magnitude, so that pixel values stayed well within saturation levels. The optical CCD exposures were binned 2×2 to facilitate rapid readout.

Flat-field observations were acquired at all three observatories using either twilight sky flats, dome flats, or a series of night-sky exposures that incorporated pointing offsets to allow removal of stars via a median filter. Standard, dark-field corrections were also applied.

### 2.2 PHOTOMETRY

Subsequent to dark current subtraction and division by a flat-field frame, we performed aperture photometry on the target star and the comparison stars using standard and custom IDL routines. In all cases, except for the McMath-Pierce telescope observations, between 2 and 8 stars of similar magnitude to the target star were used for comparison. This allowed for inter-comparison between these stars to make sure no variability was detected in them. Due to the characteristics of the heliostat (image rotation) and the small FOV available at the solar telescope only the comparison star nearest to the target star was of use. The apertures selected to measure the stars and background varied depending on the degree of defocus and seeing conditions for each observing session. These were chosen such that they minimized the scatter on the final lightcurve. The defocus on the 2.1-m telescope in particular was sensitive to changes throughout the night, due to mechanical flexure and temperature variations. We eventually learned to actively adjust the defocus setting gradually during the observations, so as to maintain image stability. For those data that exhibit variable defocus, we adjusted the numerical apertures accordingly in the data analysis process. Best results were also obtained by averaging the ratios of the target star to each comparison star. This produced similar or smaller scattering than the method of ratioing the target star to the sum of all the comparison stars. In most cases the comparison stars were of similar brightness (± ~1.5 magnitudes) as the target star. Uncertainties for each photometric point were estimated as the standard deviation of the ratio to the individual comparison stars, divided by the square root of their number (error of the mean). In all cases the observed scatter in the photometry was larger than the estimated formal uncertainties, suggesting that the errors in our photometry procedure may be underestimated. This may be due perhaps to inadequate estimation of the uncertainty in the background level in the IDL/ASTROLIB/APER photometry algorithm we use.

After normalizing the target star to the comparison stars some gradual variations as a function of time were found in some instances. In the case of the optical observations the variation was removed by using a linear airmass-dependent function fit to the baseline before and after the transit. Most transits have at least one hour's worth of baseline observations before the transit ingress and after the transit egress for this purpose. However, the near-IR observations exhibited a more complex baseline variation that could not be attributed to simple airmass-dependent comparison star differential extinction. These are most likely due to telluric waver vapor absorption variations, and/or to other instrumental effects. For these cases polynomial functions of order 2-5 were used to fit the baseline photometry. Most of the near-IR lightcurves included longer pre-ingress and post-egress observations which allowed for improved baseline fits.

Figures 1a and 1b show the near-IR transits observed with the 2.1-m telescope, while Fig. 2a (left panel) shows the four transits observed with the McMath-Pierce Telescope, and Figs. 2a and 2b show the optical transits observed with the Visitor Center Telescope. During the observing runs other transits were recorded as well, but they are either incomplete (show the ingress or egress only) or suffered from clouds, and are therefore of limited use and not included in this work.

## 3. MODELING

In order to fit the observed transit lightcurves we first created initial standard model lightcurves. These were constructed numerically as a tile-the-star procedure using the Binary Maker II software (Bradstreet 2005). The initial system parameters used were obtained from the latest literature available. Linear limb-darkening function coefficients were taken from Claret (2000). For most cases the initial model lightcurves yielded very good agreement with the observed ones and only small adjustments to the duration and depth of the model transits were necessary to optimize the fits. This was done by applying small (< 2% on average) multiplicative factors to both the depth and duration of the model transit. In most cases these small corrections fall within the published uncertainties. However, a few of the transits exhibited larger model deviations and further study was required. These are explained for the individual systems in section 4.2.

Table 1 presents all the observed transits and the principal lightcurve parameters (mid-transit time, depth, and duration) derived from fitting the models as explained above. Because the scatter of the photometry is larger than the formal errors suggest, for each lightcurve we used the scatter to estimate the uncertainties.

In order to derive improved orbital periods for the transiting systems we utilized the transit timings reported in the published literature, including the observations from this paper, and we implemented a least-squares linear fit to the data, weighting the individual transit times by their uncertainties. When relevant we have converted reported Heliocentric Julian Dates (HJD) to Barycentric Julian Dates (BJD) (Eastman, Siverd & Gaudi 2010) and have used the Dynamical Time-based system (BJD_TDB) instead of the Coordinated Universal Time-based system (BJD_UTC). The difference between heliocentric and barycentric times can be up to about four seconds, but most often it is less than this value and well within the individual timing uncertainties reported, so it has limited effect on the derived periods. However, the difference between UTC-based and TDB-based timings is a systematic offset which depends on recent additions of leap seconds to UTC. To convert BJD_TDB to BJD_UTC subtract 0.000766 days for transits observed after 1 January 2009 (JD 2,454,832.5), 0.000754 days from transits between 1 January 2006 (JD 2,453,736.5) and 1 January 2009, and 0.000743 days from transits between 1 January 1999 (JD 2,451,179.5), and 1 January 2006. The resulting system periods are presented in Table 2 along with the number of mid-transit timings used and the time span between the first and last observation reported. The reference epoch presented (JD0) is the result from the fit and generally corresponds with the first reported transit found in the literature.

For some of the best lightcurves obtained with the KPNO 2.1m telescope (J- and JH-band) we proceeded to fit theoretical transit curves to the transit data using the Transit Analysis Package (TAP) which uses Bayesian probability distributions with Markov Chain Monte Carlo (MCMC) techniques (Gazak, Tonry & Johnson, 2011). For this modeling we fixed the period and the eccentricity of each system, and used quadratic limb darkening coefficients for the stars from Claret (2000). We ran 5 chains with $10^5$ samples for each system. During testing of the software we found out that extending the modeling to $10^6$ samples did not yield significant improvement on the fit, not justifying the larger computing requirements. There was also no significant difference in using either a single linear (u) or two quadratic (a & b) coefficients to describe the stellar limb-darkening, in accordance with Southworth (2008) who found out that linear limb darkening was adequate for the analysis of high quality ground-based data. The derived lightcurve parameters of interest are reported in Table 3: the orbital inclination (i), the planet-to-star radius ratio ($R_p/R_*$) and the scaled semimajor axis ($a/R_*$).

## 4. RESULTS AND DISCUSSION

We here discuss the results of modeling our 2.1-meter J-band transit lightcurves. Sec. 4.1 discusses the overall comparison to planetary radii derived at optical wavelengths and Sec. 4.2 discusses individual systems. While we generally find good agreement with optical results, some of our J-band lightcurves do suggest variations from the published models which require further study. However, some of the systems in which we detect differences in the inclination and the scaled semimajor axis may have degenerate solutions since these two parameters trade off against each other, particularly for IR data where the stellar disk has little limb darkening. All the observations presented in this paper will be made available to the public through the NASA/IPAC/NExSci Star and Exoplanet Database (NStED) (or are available from the main author upon request). Both raw data and fitted data will be made available since we realize that the system parameters are sensitive to the baseline modeling method chosen. The simple method selected in this paper was primary aimed at establishing an accurate mid-transit time.

## 4.1 COMPARISON WITH OPTICAL PLANETARY RADII

Stellar limb darkening is much less prominent in the near-infrared as compared to optical wavelengths, so the depth of an infrared transit is closely proportional to $(R_p/R_*)^2$. One significant source of uncertainty in planetary radii from IR transit curves is the definition of the photometric baseline, which can vary in broad-band IR observations due to differences in spectral type between the target star and the comparison stars (e.g., Deming et al. 2011). However, since this source of error should vary independently from one transiting system to another, it represents random - not systematic - error when comparing an ensemble of planetary radii at optical and IR wavelengths. Figure 3 plots $R_p/R_*$ from our 2.1-m J-band transits (Table 3) against published optical radii. Potential sources of systematic differences in near-IR versus optical planetary radii include errors in stellar limb darkening, stellar activity (which affects the near-IR data less than the optical), as well as potential real differences in planetary radii with wavelength. The latter could be produced, for example, by high altitude haze (Sing et al. 2011) which increases transit radii at the shortest wavelengths.

The large stellar photon flux available to optical observers usually produces significantly smaller random errors than for our J-band transits. Consequently, we make the approximation that all of the error lies in our J-band values for $R_p/R_*$, and we perform an error-weighted linear least squares fit, shown by the solid line in Fig. 3. This fit yields:

$$(R_p/R_*)_J = 0.0017 + 0.979\,(R_p/R_*)_{optical}$$

The 1-sigma error on the slope of this relation is ±0.025, so the best-fit relation differs from the null hypothesis by less than 1-sigma. If we omit the three seemingly discrepant systems having the largest J-band errors (HAT-P-4, WASP-2 and WASP-48), the slope becomes 0.9765 ± 0.025, still better than 1-sigma agreement with the null hypothesis. We conclude that the optical radii are not likely to be affected at the approximately 5-percent level (>2-sigma) by errors in optical limb darkening. As for real differences in exoplanetary radii as a function of wavelength, we note that the variation in $R_p/R_*$ seen by Sing et al. (2011) for HD189733b is a 1.3-percent effect from the near-IR to 400 nm. That variation remains beyond the sensitivity of our statistical relation, especially since many of the optical radii apply to wavelengths longward of 400 nm (e.g., R-, I-, or z'-band). Nevertheless, we are encouraged by the prospect that further improvements in our J-band photometry (higher precision, and more transits per system), combined with improved ground-based precision in the bluer near 400 nm, could potentially provide ground-based statistical detection of haze in giant exoplanetary atmospheres, whereas such detections have so far required space-borne observations.

## 4.2 DISCUSSION OF INDIVIDUAL SYSTEMS

### 4.2.1 COROT-1

The brightness of the CoRoT-1 (V=13.6, K=12.1) star system was beyond our practical observing limit and thus resulted in a lightcurve with considerable scatter (Fig. 1a), and thus no attempt was made to derive the system parameters. It is however still of use in confirming its previously published period because of the extended time coverage. For this we used the original discovery transits of Barge et al. (2008) as timed by Bean (2009), a low-precision prediscovery transit of Rauer et al. (2010), and a high-precision transit observed by Gillon et al. (2009b) at the VLT. The Gillon et al. ephemeris predicts a midtransit time for our observation of about BJD 2,455,162.91696 ± 0.00024. Our observed midtransit time of 2,455,162.91621 ± 0.00059 falls somewhat short of the prediction, but our relatively large uncertainty still allows for a match. Bean (2009) and Csizmadia et al. (2010) had found no significant periodic timing variations with a period shorter than the original observational window of 55 days. The Gillon et al. (2009b) observation and the present one extend this time period and are also consistent with a fixed period for the planet. Including our observation we find a period for CoRoT-1b of 1.5089682 ± 0.0000005 days.

### 4.2.2 COROT-2

CoRoT-2 is a transiting system which exhibits clear evidence of starspots that have been used to estimate the rotation period of the star (Lanza et al. 2009). Alonso et al. (2008) in their discovery paper present an ephemeris which summarizes the 78 transits observed by the CoRoT mission. We combined this ephemeris with prediscovery transits reported by Rauer et al. (2010), one ephemeris reported by Vereš et al. (2009) and our current measurement (Fig. 2a) to calculate the period of the system. Our result of P=1.7429971 ± 0.0000011 days agrees with the original period of 1.7429964 ± 0.0000017 days reported by Alonso et al. (2008) and further reduces the uncertainty due to the longer time baseline. However, the earliest prediscovery transit of Rauer et al. (2010) exhibits a large deviation (O-C ~ 23.3 minutes) from its predicted transit time, well outside its own 3-sigma uncertainty range, and is thus considered suspect. Eliminating this single anomalous observation results in an alternate preferred solution of P=1.7429981±0.0000011 days.

### 4.2.3 GJ 1214

The first two of the observations presented here have already been analyzed and discussed in Sada et al. (2010). See this reference for a more thorough modeling analysis of this system. However, since then other transits have been reported. Here we assemble the original midtransit observations of Charbonneau et al. (2010), as re-evaluated in Berta et al. (2011), other observations also reported in Berta et al. (2011) including two high-precision VLT transits, and those of Sada et al. (2010). We also include in the period solution 12 new full transits (ingress and egress recorded) presented by Carter et al. (2011), three transits reported by Kundurthy et al. (2011) (their best result: chain003a), four near-infrared transits observed from Hawaii (Croll et al. 2011), plus two recent unreported transit we observed simultaneously at KPNO using the 2.1m telescope (J-band, Fig. 1a) and the VC telescope (z'-band, Fig. 2a). From these 33 transits we derive a period of P=1.5804048±0.0000002 days, in complete agreement with other recent calculations and, within observational uncertainties, with no evidence of variation during the first two observing seasons. Modeling of the system parameters from the first 2.1m lightcurve, confirming the reported planet radius, is also described in Sada et al. (2010). We have not yet attempted to fit the second high-precision lightcurve, pending further observations of the system.

4.2.4 HAT-P-1

Two HAT-P-1 transits were observed with the NAC through an H-band filter on the NSO/KPNO McMath-Pierce Solar Telescope on 2007 October 08 and October 17 (Fig. 2a). These were particularly difficult observations because the image field rotation inherent to a heliostat slowly changed the reflecting mirror surface areas throughout the long observing period. Because of this, only one star was of use as a comparison source. Fortunately HAT-P-1 has a close companion. Even so, on the night of 2007 October 17 high winds hitting the main heliostat mirror introduced severe noise on the data. We used our data along with the low-precision Bakos et al. (2007) discovery paper observation, the midtransit times of Winn et al. (2007b) as corrected in Winn et al. (2008), and the Johnson et al. (2008) reported transits to derive a period of 4.4653054±0.0000069 days. In this calculation we did not include our 2007 October 17 observation since it deviated more than 3-sigma from its predicted value, probably as a result of the severe wind problem.

4.2.5 HAT-P-3

We observed two transits of HAT-P-3 on 2009 May 15 and 2010 May 27. Each transit was observed with the KPNO 2.1m telescope with a JH filter (Figs 1a) and also with the KPNO VC telescope through a z' filter the first night and a B filter on the second (Fig. 2a). We combined our derived midtransit timings with those of Torres et al. (2007), Gibson et al. (2010), Chan et al. (2011) and Nascimbeni et al. (2011) to derive a period of 2.8997382±0.0000009 days, in agreement with recent calculations but with improved uncertainty. The 2009 May 15 lightcurve in particular exhibits a gap in the data during egress and was not modeled. The 2010 May 27 lightcurve on the other hand yields a model with slightly lower orbital inclination (~$85.7^{o}$±$0.55^{o}$ vs ~$87.1^{o}$±$0.55^{o}$) than the one reported in Chan et al. (2011) and Torres et al. (2008). We also obtain a smaller scaled semimajor axis (~9.2±0.5 vs ~10.4±0.5), which combined probably explains the slightly shorter duration (~5 min) observed for this particular transit compared with Chan et al. (2011).

4.2.6 HAT-P-4

We observed one transit for HAT-P-4 on 2011 May 22 with the KPNO 2.1m telescope using a J-band filter (Fig. 1a). We combined our midtransit timing with the two found in the discovery paper (Kovács et al. 2007), one reported observation by Winn et al. (2010), and ten additional EPOXI observations (Christiansen et al. 2011) to derive a period of 3.0565254±0.0000012 days. This differs from 3.0565114±0.0000028 days reported by Christiansen. In this particular case we can trace the difference to a registered discontinuity in our data (possible due to a temporary gain fluctuation in the camera amplifiers) just at egress that had to be corrected manually. The depth of the transit is so shallow that any variation at the ingress/egress portions of the lightcurve is critical in determining the transit duration. We attempted to compensate empirically for it by matching the comparison star brightness after egress. However, this makes our egress portion of the lightcurve suspect and could account for our unusual short transit duration and shift of the midtransit time. For this system we also derive a smaller orbital inclination (~86.0$^{o}$ vs ~89.8$^{o}$) and scaled semimajor axis (~5.6 vs ~6.0) from those reported by Torres et al. (2008) and Christiansen et al. (2011) which is also reflected by the observed smaller (by ~10 min) transit duration.

#### 4.2.7 HAT-P-6

We observed one transit of HAT-P-6 on 2009 November 25 with the KPNO 2.1m telescope through a J-band filter (Fig. 1a). We have combined our midtransit timing with those of the discovery paper (Noyes et al. 2008), as reevaluated in Szabó et al. (2010), and a newer transit also reported by Szabó et al. (2010), to derive an improved period of 3.8530018±0.0000015 days. The only reported model parameters for this system correspond to the original discovery paper by Noyes et al. (2008). Our results do vary slightly from those published initially and may be an alternative, though additional observations of this system are needed to improve on our uncertainties.

#### 4.2.8 HAT-P-11

We observed one transit of HAT-P-11 with both the KPNO 2.1m telescope (J-band) and VC telescope (B-band) on 2010 June 01 (Figs. 1a and 2a). The depth of this transit is rather shallow but well defined by the good observing conditions at the time. These observations are detailed further in Deming et al. (2011). In addition we also observed a transit of HAT-P-11 with the VC telescope (B-band again) on 2011 May 14. We used our derived timings along with those reported by Bakos et al. (2010b), Ditmann et al. (2009), Hirano et al. (2011), and the Kepler mid-transit observations analyzed in Deming et al. (2011) to derive an improved period of 4.8878056±0.0000015 days. Although our reported mid-transit timings for June 2010 may be suspect (see Deming et al. 2011), they have little weight against the high-quality Kepler data. Of particular interest is the VC telescope transit observed on 2011 May 14 (Fig. 2a). We chose to observe through the B-filter so as to obtain greater sensitivity to possible stellar activity due to the shorter effective wavelength of the B filter. Our intent was to observe possible star spots as the planet, which crosses the stellar disk nearly perpendicular to its equator, covered the latitudes of interest on the surface of this known active star (see Deming et al. 2011, and Sanchis-Ojeda et al. 2011, for further details). From careful observation of the lightcurve we do see a definite feature at ~+0.019 days which corresponds with one of the locations where starspots were seen in the Kepler data. This confirms that the latitudes of stellar activity on this star can be monitored with ground-based observations using short-wavelength filters. Even after the end of the Kepler mission, ground-based observations of the transits can be used to monitor the evolution of magnetic activity on this active star (Deming et al. 2011).

### 4.2.9 HAT-P-12

The only other midtransit ephemeris for HAT-P-12 available in the literature corresponds to the discovery paper by Hartman et al. (2009). Combined with our 2.1m telescope J-band observation presented here (Fig. 1a) we obtain an improved period of 3.2130553 ± 0.0000010 days. More transit timings need to be reported in order to further improve the period of this system. Our derived model parameters from Table 3 are in agreement with those reported in Hartman et al. (2009) although our uncertainties are larger.

### 4.2.10 HAT-P-27 / WASP-40

We combined our single recent observation of this transiting system (KPNO 2.1m telescope, J-band, Fig. 1a) with the epoch reported in the discovery papers by Béky et al. (2001) and Anderson et al. (2011) to derive an improved period of 3.0395824 ± 0.0000035 days. Our derived model parameters from Table 3 are in agreement with those reported in the discovery papers although our uncertainties are larger. Anderson et al. (2011) report a 40% probability that the transits occur in a grazing configuration. The minimal limb darkening in the near-IR normally yields transits that are quite flat-bottomed. In contrast to this normal behavior, the noticeable roundness of our J-band transit curve (Fig. 1a) confirms that the transit is grazing.

### 4.2.11 HAT-P-32

One ephemeris for this system is reported in the literature resulting from the observation of several transits in the discovery paper (Hartman et al. 2011). We combined our three KPNO 2011 observations (two 2.1m telescope J-band transits on October 09 and 11 – Fig. 1a, and a VCT z'-band one on October 11 – Fig. 2a) to derive a period of 2.1500103 ± 0.0000003 days with improved uncertainty.

Our simultaneous observations of 2011 October 11 through a z'-band and a J-band filters both exhibit a small brightness increase just after midtransit that is not evident in the J-band observation from the previous planetary transit observed two days before through a J-band filter, and it is probably associated with starspot activity. The model results reported in Table 3 correspond only to modeling the 2011 October 09 lightcurve since it had a higher S/N ratio and did not have the starspot. Our resulting model inclination and scaled stellar radius agree with those reported by Hartmann et al. (2011), but our planet-to-star size ratio seems to be a slightly larger.

### 4.2.12 HD 17156

HD 17156 is a system with a relatively long orbital period (~21.2 days) and thus transit opportunities are infrequent and observations are valuable. We registered one transit of this system with the KPNO VC telescope on 2009 Nov. 24 (Fig. 2a). This star has a high northern declination, beyond the limit of the 2.1m telescope, and reachable only using the german equatorial mount of the VC telescope. We also gather the first reported transit observation by Barbieri et al. (2007), the midtransits reported by Irwin et al. (2008), Narita et al. (2008), Gillon et al. (2008), Winn et al. (2009) and the high-quality HST observations analyzed in Nutzman et al. (2011) to report a period of 21.216384±0.000016 days for the system. The solution is dominated by the three high-quality HST observations and our result deviate by about 150 seconds from the prediction based on the derived ephemeris. However, this difference is still within our measurement uncertainty.

4.2.13 HD189733

This is a well-observed transiting system with ample reported transits. We observed HD 189733 with the 2.1m telescope (J-band, Fig. 1a) and the VC telescope (Fig. 2a) on two different occasions. The latest VC telescope observation is of particular interest since, due to a filter wheel error, we observed it through an Hα filter. Careful analysis of the observed lightcurve reveals variations from a symmetric lightcurve due to chromospheric stellar features as the planet transits the disk of the active star.

For this system we obtained all ground-based (Bouchy et al. 2005, Bakos et al. 2006, Winn et al. 2007b, Hrudková et al. 2010) and spacecraft (Pont et al. 2007, Knutson et al. 2007, Miller-Ricci et al. 2008, Knutson et al. 2009, Agol et al. 2010) midtransit times reported in the literature to derive a period of 2.2185754 ± 0.0000001 days, consistent with the latest Agol et al. (2010) estimate based solely on Spitzer observations.

This is an often-studied bright system that has been observed from space and its lightcurve parameters are well constrained. Our single lightcurve observation cannot improve on those results, but it was of particular help in refining the data analysis and modeling techniques used throughout this work. Within our much larger uncertainties, our lightcurve model parameters correspond with those of the literature.

4.2.14 Qatar-1

Our single KPNO 2.1m J-band observation of this system (Fig. 1a) was combined with the discovery article ephemeris (Alsubai et al. 2010) to derive an improved period of 1.4200227 ± 0.0000012 days. The results of our modeling are in fair agreement with those reported by Alsubai et al. (2010) assuming a circular orbit.

4.2.15 TrES-1

There are few midtransit times reported in the professional literature for this system despite this being one of the earliest exoplanet systems discovered and announced. We managed to observe two consecutive transits in 2007 using the NAC array at the KPNO NSO McMath-Pierce Solar Telescope (H-band, left panel Fig. 2a). We only used one comparison star in the photometry because of problems of image field rotation due to the design nature of the heliostat telescope used. Fortunately this system has a close comparison star of similar magnitude that yielded usable lightcurves. Using our two observations and those reported in the literature (Alonso et al. 2004, Charbonneau et al. 2005, Narita et al., 2007, Winn et al, 2007c, Hrudková et al, 2009, Raetz et al. 2009b, and Rabus et al. 2009) we obtain a period of 3.0300724±0.0000004 days, slightly improving on the Rabus et al. (2009) latest value. However, we note that a longer timeline of transits observations, like the large amateur collection found in the Exoplanet Transit Database (ETD - http://var2.astro.cz/ETD/ ), is needed and might still yield a slightly different period.

4.2.16 TrES-2

We obtained two transits of this well-observed and characterized exoplanet system in 2011 with the both KPNO telescopes on May 20 (VCT z'-band, Fig. 2b) and October 13 (2.1m J-band, Fig. 1a). Combining our midtransit times with those reported from ground-based observations (O'Donovan et al. 2006, Holman et al. 2007, Raetz et al. 2009a, Mislis & Schmitt 2009, Mislis et al. 2010, Rabus et al. 2009, Scuderi et al. 2010 & Colón et al. 2010) and Kepler (Kipping & Bakos 2011) and EPOXI (Christiansen et al. 2011) spacecraft data we obtain a period of 2.4706128±0.0000003 days. Our period matches the one derived from combining several observations from diverse sources over a long period (Christiansen et al. 2011) but is less than the period derived using short-term, but very high-quality, Kepler spacecraft observations (Kipping & Bakos 2011). Our single near-IR lightcurve modeling cannot compare with better quality data available for this well studied system. However, our results do agree very well with those published for this system, within our larger uncertainties.

4.2.17 TrES-3

We observed one transit of TrES-3, with the KPNO VC telescope (z'-band, Fig. 2b). We gathered the midtransit times reported in the literature by Sozzetti et al. (2008), which includes a reevaluation of the midtransit time from the O'Donovan et al. (2006) announcement paper, Gibson et al. (2009), Colón et al. (2010), Christiansen et al. (2011) and Woo-Lee et al. (2011) to derive a period of 1.3061865±0.0000002 days for this system. This period agrees with the one presented by Christiansen et al. (2011) and Vaňko et al. (2011), but is slightly shorter than the 1.30618700±0.00000015 day period reported by Woo-Lee et al. (2011) that includes a large number of amateur observations.

4.2.18 TrES-4

We observed one transit of TrES-4 with both the KPNO 2.1m telescope (J-band, Fig. 1b) and VC telescope (B-band, Fig. 2b) on 2010 May 30. The depth of this transit is shallow but well defined by the good observing conditions at the time. Unfortunately there was also a timing issue that night with the 2.1m telescope software and we cannot derive a trustworthy midtransit time for that observation. We therefore use our single VC telescope timing along with those reported by Sozzetti et al. (2009), which are re-analyzed original observations from Mandsushev et al. (2007), and those of Chan et al. (2011) to derive an improved period of 3.5539303±0.0000019 days. Our derived lightcurve model parameters from Table 3 are also in agreement, within our uncertainties, with those recently reported by Chan et al. (2011).

4.2.19 WASP-1

We observed WASP-1 on one occasion with the KPNO 2.1m telescope (J-band, Fig. 1b). This system has few midtransit observations reported in the literature. We used the discovery ephemeris (Collier Cameron et al. 2007) along with the midtransit reports from Charnonneau et al. (2007), Shporer et al. (2007), Wang et al. (2008) and Albrecht et al. (2011) to derive a period of 2.5199425±0.0000014 days. In our analysis we did not include the midtransit report from Szabó et al. (2010) because it deviates by over 14 minutes from the predicted time, well outside of its reported uncertainty. Thus we consider the time suspect. Our derived parameters for this system fall just within those lately reported by Simpson et al. (2011).

4.2.20 WASP-2

We observed WASP-2 on one occasion with the KPNO 2.1m telescope (J-band, Fig. 1b). This system also has few midtransit observations reported in the literature. We used these (Collier Cameron et al. 2007, Charbonneau et al. 2007, Hrudková et al. 2009 and Southworth et al. 2010) to derive a period of 2.1522213±0.0000004 days, in agreement with the report of Southworth et al. (2010) that also includes a large number of amateur observations. Southworth et al. (2010) also do a thorough job of comparing various model parameters reported in the literature, and our results agree within our larger uncertainties.

4.2.21 WASP-3

WASP-3 is a well observed system with ample mid-transit timings found in the literature. Here we gathered the earlier observations from Pollaco et al. (2008), Gibson et al. (2008), a low-precision measurement by Damasso et al. (2009), well observed events by Triapathi et al. (2010), Maciejewski et al. (2010), Littlefield (2011), and space-based observations by Christiansen et al. (2011) and combine them with our four observations (three from the VC telescope and one from the 2.1m telescope – Figs. 1b and 2b) to derive an improved period of 1.8468332±0.0000004 days. In particular Maciejewski et al. (2010) propose that the observed mid-transit time deviations from a constant period may be due to the possible presence of a ~15 earth-mass planet located close to the 2:1 mean motion resonance. The observations from Littlefield (2011) seem to support in part this conclusion. Our transits do not directly confirm these timing variations. The measured midtransit uncertainties for the lightcurves, obtained using different methodologies, accompanied with time synchronization and standardization issues, are just large enough to confuse the issue. Our modeling results also agree with those reported in Christiansen et al. (2011) from EPOXI observations.

4.2.22 WASP-6

We have combined our 2011 October 12 J-band KPNO 2.1m midtransit time (Fig. 1b) with the one reported on the discovery paper (Gillon et al. 2009a) and the observation from (Dragomir et al. 2011) to derive an improve period of 3.3609998 ± 0.0000011 days. Our model for this system agrees with the inclination and semimajor axis reported by Gillon et al. (2009a) and Dragomir et al. (2011). However, we obtain a smaller planet-to-star ratio in our near-IR observations.

4.2.23 WASP-10

WASP-10 was our first system observed transiting at both the KPNO 2.1m telescope (J-band, Fig. 1b) and the VC telescope (z'-band, Fig. 2b). Our midtransit times differ from one another primarily due to the relative scarcity of data during egress at the VC telescope, although they still overlap at the 3-sigma level. We gathered other existing midtransit times from the discovery paper by Christian et al. (2009), from the high-quality observation by Johnson et al. (2009) corrected in Johnson et al. (2010), from Dittman et al. (2010), and from Maciejewski et al. (2011a) and Maciejewski et al. (2011c). Maciejewski et al. (2011a) also reanalyzed four midtransit times presented in Krejčová et al. (2010). All these observations yield a period of 3.0927297±0.0000003 days.

Maciejewski et al. (2011a) report midtransit timing variations which can be explained by the presence of an additional planet about a tenth of the mass of Jupiter orbiting close to the outer 5:3 mean motion resonance with a period of about 5.23 days. Our single observations at epoch 129 (see their Figure 2) would yield an O-C of about +2.35 minutes from their linear ephemeris. Our observed deviations are ~+0.7±0.3 minutes for the 2.1m and ~+2.5±0.6 minutes for the VC telescope. Although the VC telescope observation might agree with the prediction, we have a larger confidence in the mid-transit time derived from the 2.1m telescope lightcurve. Thus we are unable to either support or deny the claim of a second planet in the system. In fact, our higher quality 2.1m observation matches the Maciejewski et al. (2011c) four high-quality observations very well if we use the period derived here. An alternative, and more general, explanation could be that most midtransit times are susceptible to larger variations than their formal uncertainties reflect because of either incomplete lightcurves and baseline trends due to stellar variability and/or differential extinction of the field comparison stars which have not been identified and accounted for properly.

There has been some discussion in the literature concerning the radius of WASP-10b (Johnson et al. 2009, Dittman et al. 2010). Our J-band transit result yields a planetary to stellar radius ratio in close agreement with Johnson et al. (2009).

4.2.24 WASP-11/HAT-P-10

We observed four transits of this system on 2009 November 26 with the KPNO 2.1m telescope, on 2010 November 07 with the KPNO VC telescope, and on 2011 October 08 with both telescopes (See Figs. 1b and 2b). We combined our midtransit timings with those reported on both discovery papers (West et al. 2009 and Bakos et al. 2010) to derive an improved period of 3.7224793±0.0000007 days.

Our higher quality J-band lightcurve from 2009 November 26 shows an overall depth of 1.8±0.2%, which is smaller than the reported depth for this system. In addition, the shape of the bottom of the transit is not smooth and does not match the fitted model well. Specifically, our observed depth seems to be greater just before the beginning of transit egress, which suggests that the real depth is indeed larger and that spots could be present on the stellar surface. To test this theory we ran several models in which we adopted the best system parameters and then proceeded to spot the stellar surface during the first part of the transit. The result is presented in Fig. 4. The solid line represents the standard literature model while the dotted line is our best overall fit to the observed depth of the lightcurve. The dashed line is the standard literature model with part of the stellar surface covered by starspots with temperatures ~6% lower than the effective temperature of the star (4625 K vs 4920 K). Although we point out that this solution is not unique and no effort was made to find the best fit possible. Note how the middle part of the transit and the brightness drop at phase ~0.025 is better matched by the spotted model. Spectroscopy of the star by Knutson et al. (2010) showed that it exhibited the $5^{th}$ most intense K-line emission of 50 stars in their sample, a clear indicator of stellar activity. Further evidence that the star may exhibit starspots is found in the Exoplanet Transit Database (Poddany, Brat & Pejcha 2010), a compilation of professional and amateur planetary transit lightcurves. Reported observations for this system show variations in recorded transit depths that go from about 16 to 26 milimagnitudes, Our observed transit corresponds to their epoch #108, close in time to a reported depth of ~16 milimagnitudes on transit #106. Our KPNO VC telescope z'-band transit one observing season later shows a deeper 2.1% transit, although with higher uncertainty; and both transit lightcurves from 2011 October 08 show a depth of ~1.9%. From this we conclude that the star exhibits starspot activity that affects the transits and thus the derived model parameters. On Table 3 we report our model results from both of our J-band lightcurves which agree very well with each other, and fall somewhere in between those reported by both discovery papers.

4.2.25 WASP-12

We observed one transit of WASP-12 with the KPNO VC telescope (Fig. 2b, z'-band). We have combined our observation with those published by Hebb et al. (2009), Chan et al. (2011) and Maciejewski et al. (2011b) to derive a period of 1.0914224±0.0000003 days in agreement with the last two publications.

4.2.26 WASP-24

We observed one transit of WASP-24 with the KPNO 2.1m telescope (J-band, Fig. 1b) and combine our midtransit time with those reported in the discovery paper by Street et al. (2010) to derive a period of 2.3412162 ± 0.0000014 days. Simpson et al. (2011) publishes model parameters that match those of Street et al. (2010). Our derived results seem to deviate significantly, showing a larger system inclination (~86.4$^{o}$ vs. 83.6$^{o}$) compensated by a larger scaled semimajor axis (~7.1 vs. 6.0), although our larger uncertainties might still allow for a match. The planet size is similar in both studies.

4.2.27 WASP-32

We observed one transit of WASP-32 through a J-band filter with the KPNO 2.1m telescope on 2011 October 15 (Fig. 1b). We combine our result with the ephemeris presented in the discovery paper (Maxted et al. 2010) to derive an improved period of 2.7186591 ± 0.0000024 days. Our model results agree with those reported by Maxted et al. (2010) with respect to the inclination and scaled semimajor axis of the system. Our derived planet-to-star radius ratio (0.1030) is smaller than their reported radius ratio (0.1113) by about 3-sigma.

4.2.28 WASP-33

WASP-33 is the hottest known hot Jupiter (Smith et al. 2011) closely orbiting a bright delta Scuti variable host star (Herrero et al. 2011). Only two epochs are reported in the literature: in the discovery paper (Collier Cameron et al. 2010) and by Smith et al. (2011). We combine these midtransit times with our observations: a z'-band lightcurve from the KPNO VC telescope on 2010 November 03 and two other lightcurves obtained on 2011 October 13 (a J-band one at the 2.1m telescope, and a Hydrogen Alpha observations with the VCT), shown in Figs. 1b and 2b, to derive a period of 1.2198721±0.0000003 days.

Both observations obtained on 2011 October 13 clearly show short-period variability that interferes with a clean determination of the lightcurve parameters on Table 1 and the near-IR model parameters of Table 3. This is evidenced as an offset of the baseline before ingress compared with the egress baseline, and there is also an increase in brightness affecting the first half of the transit depth. These effects are enhanced on the shorter wavelength and narrower bandpass Hydrogen Alpha lightcurve that shows the transit of the planet against the chromosphere of the star compared with the J-band one which shows the same transit against a smoother and less limb-darkened stellar disk. Simultaneous observations of the same transit at different wavelengths can help limit stellar surface characteristics (like wavelength-dependent brightness amplitude variability in this case). Because the above mentioned stellar variability was not accounted for, the resulting model parameters in Table 3 are suspect.

4.2.29 WASP-48

There is only one discovery epoch reported for this system (Enoch et al. 2011). We combine this midtransit time with our KPNO 2.1m J-band observation (Fig. 1b) to derive a period of 2.1436283±0.0000041 days. For this system we also derive parameters that seem to deviate from the ones reported in the discovery paper by Enoch et al. (2011). Our inclination seems to be larger (~85.1$^o$ vs. ~80.1$^o$) also followed by a larger scaled semimajor axis (~5.4 vs. ~4.2). However, the planet size seems to be the same between both studies.

4.2.30 WASP-50

We combined our two consecutive 2011 observations of this system from KPNO (z'-band with the VCT on October 15 – Fig. 2b, and J-band with the 2.1m on October 17 – Fig. 1b) with the reported ephemeris on the discovery article (Gillon et al. 2011) to derive an improved period of 1.9550905 ± 0.0000022 days thanks to lengthening the time baseline of observations. Our modeling results are also consistent with those reported by Gillon et al. (2011).

4.2.31 XO-1

We observed one transit of this system with the KPNO VC telescope (z'-band, Fig. 2b). We combined our midtransit timing with those reported by McCullough et al. (2006), Holman et al. (2006), Wilson et al. (2006), Vaňko et al. (2009), Raetz et al. (2009b), Cáceres et al. (2009), and the HST observations by Burke et al. (2010) to derive a period of 3.9415052±0.0000008 days in agreement with the latest report by Burke et al. (2010).

4.2.32 XO-5

We observed one transit of XO-5 with the KPNO 2.1m telescope (J-band, Fig. 1b). We combined our midtransit timing with the low precision observations reported by Burke et al. (2008), the better precision observations of Pál et al. (2009), and two additional observations from Maciejewski et al. (2011c) to derive a period of 4.1877544±0.0000016 days. The system parameters also fall within previously reported values.

TABLE 1

EXTRASOLAR PLANET TRANSIT INFORMATION

| Name | Date | Tel. (1) | Filter (2) | # C.S. (3) | Base. Fit (4) | Mid-Transit Time (5) (BJD_TDB) | Duration (min) | Depth (%) |
|---|---|---|---|---|---|---|---|---|
| CoRoT-1 | Nov. 27, 2009 | 2.1m | JH | 3 | P5 | 5162.91698 ± 0.00059 | 147.2 ± 2.4 | 1.96 ± 0.65 |
| CoRoT-2 | May 09, 2009 | VC | z’ | 8 | X | 4960.88065 ± 0.00044 | 140.4 ± 1.8 | 3.23 ± 0.51 |
| GJ 1214 | May 29, 2010 | 2.1m | J | 7 | P5 | 5345.82204 ± 0.00011 | 52.7 ± 0.5 | 1.47 ± 0.16 |
| | May 29, 2010 | VC | B | 8 | X | 5345.82221 ± 0.00032 | 55.1 ± 1.3 | 1.74 ± 0.39 |
| | May 18, 2011 | VC | z’ | 4 | --- | 5699.83288 ± 0.00034 | 54.6 ± 1.4 | 1.42 ± 0.37 |
| | May 18, 2011 | 2.1m | J | 8 | P1 | 5699.83283 ± 0.00014 | 52.7 ± 0.6 | 1.44 ± 0.16 |
| HAT-P-1 | Oct. 08, 2007 | MP | H | 1 | P1 | 4381.81060 ± 0.00077 | 174.8 ± 3.1 | 1.68 ± 0.40 |
| | Oct. 17, 2007 | MP | H | 1 | P1 | 4390.74563 ± 0.00139(6) | 181.5 ± 5.7 | 1.47 ± 0.66 |
| HAT-P-3 | May 15, 2009 | 2.1m | JH | 3 | P3 | 4966.89186 ± 0.00025 | 123.6 ± 1.0 | 1.27 ± 0.20 |
| | May 15, 2009 | VC | z’ | 4 | X | 4966.89323 ± 0.00047 | 115.4 ± 1.9 | 1.36 ± 0.28 |
| | May 27, 2010 | 2.1m | JH | 3 | P5 | 5343.85846 ± 0.00021 | 119.5 ± 0.9 | 1.25 ± 0.23 |
| | May 27, 2010 | VC | B | 6 | X | 5343.85867 ± 0.00048 | 129.1 ± 2.0 | 1.35 ± 0.28 |
| HAT-P-4 | May 22, 2011 | 2.1m | J | 2 | P1(7) | 5703.77929 ± 0.00060 | 244.2 ± 2.5 | 0.73 ± 0.21 |
| HAT-P-6 | Nov. 25, 2009 | 2.1m | J | 8 | P5 | 5160.75292 ± 0.00034 | 204.4 ± 1.4 | 0.96 ± 0.16 |
| HAT-P-11 | Jun. 01, 2010 | 2.1m | J | 6 | P2 | 5348.83923 ± 0.00027(6) | 146.4 ± 1.1 | 0.45 ± 0.11 |
| | Jun. 01, 2010 | VC | B | 2 | --- | 5348.83574 ± 0.00055 | 145.6 ± 2.3 | 0.50 ± 0.23 |
| | May 14, 2011 | VC | B | 5 | X | 5695.87244 ± 0.00105 | 158.7 ± 4.3 | 0.50 ± 0.21 |
| HAT-P-12 | May 31, 2010 | 2.1m | J | 5 | P5 | 5347.76929 ± 0.00021 | 138.3 ± 0.9 | 2.27 ± 0.28 |
| HAT-P-27/W40 | May 21, 2011 | 2.1m | J | 7 | P2 | 5702.74876 ± 0.00039 | 97.5 ± 1.6 | 1.28 ± 0.15 |
| HAT-P-32 | Oct. 09, 2011 | 2.1m | J | 7 | P1 | 5843.75341 ± 0.00019 | 187.1 ± 0.8 | 2.59 ± 0.20 |
| | Oct. 11, 2011 | 2.1m | J | 6 | P1 | 5845.90287 ± 0.00024 | 185.5 ± 1.0 | 2.21 ± 0.22 |
| | Oct. 11, 2011 | VC | z' | 5 | X | 5845.90314 ± 0.00040 | 183.2 ± 1.6 | 2.39 ± 0.21 |
| HD 17156 | Nov. 24, 2009 | VC | z' | 4 | --- | 5159.84014 ± 0.00068 | 182.3 ± 2.7 | 0.72 ± 0.30 |
| HD 189733 | Oct. 19, 2008 | 2.1m | J | 5 | P2 | 4758.64910 ± 0.00011 | 109.8 ± 0.4 | 2.49 ± 0.16 |
| | May 13, 2011 | VC | Ha | 2 | X | 5694.88813 ± 0.00025 | 108.5 ± 1.0 | 2.68 ± 0.35 |
| Qatar-1 | Oct. 16, 2011 | 2.1m | J | 4 | P3 | 5850.69628 ± 0.00019 | 101.3 ± 0.8 | 2.46 ± 0.19 |
| TrES-1 | Oct. 10, 2007 | MP | H | 1 | X | 4383.68588 ± 0.00072 | 150.3 ± 2.9 | 1.85 ± 0.47 |
| | Oct. 13, 2007 | MP | H | 1 | --- | 4386.71733 ± 0.00078 | 147.5 ± 3.2 | 1.58 ± 0.53 |
| TrES-2 | May 20, 2011 | VC | z' | 5 | X | 5701.88715 ± 0.00068 | 110.4 ± 2.8 | 1.43 ± 0.28 |
| | Oct. 13, 2011 | 2.1m | J | 8 | --- | 5847.65447 ± 0.00029 | 110.2 ± 1.2 | 1.63 ± 0.19 |
| TrES-3 | May 06, 2009 | VC | z' | 7 | X | 4957.86698 ± 0.00048 | 77.9 ± 1.9 | 2.49 ± 0.48 |
| TrES-4 | May 30, 2010 | 2.1m | J | 5 | P3 | 5346.83509 ± 0.00061(6) | 217.1 ± 2.5 | 0.91 ± 0.17 |
| | May 30, 2010 | VC | B | 4 | X | 5346.84102 ± 0.00051 | 215.8 ± 2.1 | 0.89 ± 0.18 |
| WASP-1 | Oct. 22, 2008 | 2.1m | J | 8 | P2 | 4761.73558 ± 0.00033 | 226.0 ± 1.4 | 1.24 ± 0.23 |
| WASP-2 | Oct. 18, 2008 | 2.1m | J | 7 | P3 | 4757.70492 ± 0.00032 | 109.8 ± 1.3 | 1.41 ± 0.17 |
| WASP-3 | May 12, 2009 | VC | z' | 8 | X | 4963.84563 ± 0.00055 | 160.9 ± 2.3 | 1.09 ± 0.26 |
| | Jun. 02, 2010 | 2.1m | J | 5 | P2 | 5349.83457 ± 0.00039 | 161.6 ± 1.6 | 1.01 ± 0.14 |
| | Jun. 02, 2010 | VC | z' | 4 | X | 5349.83182 ± 0.00039 | 156.8 ± 1.6 | 1.15 ± 0.21 |
| | May 17, 2011 | VC | z' | 6 | P1 | 5698.88358 ± 0.00060 | 150.4 ± 2.4 | 1.09 ± 0.29 |
| WASP-6 | Oct. 12, 2011 | 2.1m | J | 7 | P2 | 5846.72540 ± 0.00045 | 157.8 ± 1.8 | 2.18 ± 0.24 |
| WASP-10 | Oct. 17, 2008 | VC | z' | 6 | X | 4756.82125 ± 0.00039 | 132.5 ± 1.6 | 3.05 ± 0.26 |
| | Oct. 17, 2008 | 2.1m | J | 5 | P2 | 4756.81997 ± 0.00018 | 134.2 ± 0.7 | 2.92 ± 0.28 |
| WASP11/HP10 | Nov. 26, 2009 | 2.1m | J | 7 | P3 | 5161.71529 ± 0.00021 | 158.6 ± 0.9 | 1.78 ± 0.19 |
| | Nov. 07, 2010 | VC | z' | 5 | --- | 5507.90419 ± 0.00042 | 163.1 ± 1.7 | 2.09 ± 0.35 |
| | Oct. 08, 2011 | 2.1m | J | 6 | --- | 5842.92921 ± 0.00021 | 154.3 ± 0.8 | 1.87 ± 0.25 |
| | Oct. 08, 2011 | VC | z' | 5 | P1 | 5842.92952 ± 0.00044 | 158.1 ± 1.8 | 1.94 ± 0.24 |
| WASP-12 | Oct. 29, 2010 | VC | z’ | 8 | X | 5498.89590 ± 0.00079 | 177.2 ± 3.2 | 1.39 ± 0.45 |
| WASP-24 | May 20, 2011 | 2.1m | J | 3 | P2 | 5701.80338 ± 0.00049 | 152.6 ± 2.0 | 1.12 ± 0.31 |
| WASP-32 | Oct. 15, 2011 | 2.1m | J | 3 | P2 | 5849.75000 ± 0.00037 | 139.5 ± 1.5 | 1.16 ± 0.23 |
| WASP-33 | Nov. 03, 2010 | VC | z’ | 6 | P1 | 5503.86346 ± 0.00035 | 167.5 ± 1.4 | 1.14 ± 0.20 |
| | Oct. 13, 2011 | 2.1m | J | 4 | --- | 5847.86796 ± 0.00032 | 169.9 ± 1.3 | 0.95 ± 0.14 |
| | Oct. 13, 2011 | VC | Ha | 2 | --- | 5847.86974 ± 0.00072 | 155.3 ± 2.9 | 1.16 ± 0.32 |
| WASP-48 | May 15, 2011 | 2.1m | J | 8 | --- | 5696.81358 ± 0.00057 | 190.3 ± 2.3 | 1.07 ± 0.19 |
| WASP-50 | Oct. 15, 2011 | VC | z’ | 2 | X | 5849.92131 ± 0.00060 | 108.6 ± 2.5 | 2.11 ± 0.32 |
| | Oct. 17, 2011 | 2.1m | J | 2 | P2 | 5851.87634 ± 0.00028 | 108.8 ± 1.2 | 1.84 ± 0.19 |
| XO-1 | May 08, 2009 | VC | z’ | 2 | --- | 4959.83598 ± 0.00039 | 177.6 ± 1.6 | 2.11 ± 0.38 |
| XO-5 | Nov. 24, 2009 | 2.1m | J | 7 | P5 | 5159.89907 ± 0.00038 | 188.3 ± 1.5 | 1.19 ± 0.27 |

(1) Telescope Used: 2.1m = KPNO 2.1m telescope, VC = KPNO Visitor Center 0.5m telescope, MP = KPNO National Solar Observatory McMath-Pierce 2.0m Telescope.
(2) Filter Used: B = Johnson Blue (0.44 μm), Ha = Hydrogen Alpha (0.656 μm), z’ = Sloan DSS z’ (0.90 μm), J = mid-IR J-band (1.25 μm), H = mid-IR H-band (1.64 μm), JH = mid-IR J- and H-bands combined (1.25 and 1.64 μm)
(3) # C.S.: Number of field comparison stars used to derive the lightcurve.
(4) Base. Fit: Type of fit used on the out-of-transit baseline. Pn = polynomial of order n, X = airmass dependence, --- = none used.
(5) Barycentric Julian Date based on Dynamical Time (2,450,000+). To convert to BJD_UTC (Coordinated Universal Time) subtract 0.00075 days from the 2008 times, and 0.00077 days from the 2009 and 2010 times.
(6) Timing issue with telescope. Not trustworthy.
(7) Severe discontinuity in data after egress. Fixed manually.

TABLE 2

DERIVED SYSTEM EPHEMERIDES

| Name | Number of Observations | Time Coverage (years) | Period (days) | JD0 (BJD_TDB) (2,450,000+) |
|---|---|---|---|---|
| CoRoT-1 | 38 | 3.0 | 1.5089682 ± 0.0000005 | 4138.32807 ± 0.00006 |
| CoRoT-2 | 5 | 3.8 | 1.7429981 ± 0.0000011 | 4237.53639 ± 0.00014 |
| GJ 1214 | 33 | 2.0 | 1.5804048 ± 0.0000002 | 4964.94469 ± 0.00006 |
| HAT-P-1 | 11 | 1.1 | 4.4653054 ± 0.0000069 | 3984.39735 ± 0.00026 |
| HAT-P-3 | 19 | 3.1 | 2.8997382 ± 0.0000009 | 4218.75959 ± 0.00026 |
| HAT-P-4 | 14 | 4.0 | 3.0565254 ± 0.0000012 | 4245.81521 ± 0.00020 |
| HAT-P-6 | 4 | 1.9 | 3.8530018 ± 0.0000015 | 4035.67618 ± 0.00025 |
| HAT-P-11 | 32 | 6.8 | 4.8878056 ± 0.0000015 | 4605.89123 ± 0.00013 |
| HAT-P-12 | 2 | 2.5 | 3.2130553 ± 0.0000010 | 4419.19631 ± 0.00020 |
| HAT-P-27/WASP-40 | 3 | 1.4 | 3.0395824 ± 0.0000035 | 5186.01982 ± 0.00032 |
| HAT-P-32 | 4 | 3.9 | 2.1500103 ± 0.0000003 | 4420.44637 ± 0.00009 |
| HD 17156 | 10 | 2.2 | 21.216384 ± 0.000016 | 4353.61930 ± 0.00034 |
| HD 189733 | 59 | 5.7 | 2.2185754 ± 0.0000001 | 3629.39489 ± 0.00003 |
| Qatar-1 | 2 | 0.9 | 1.4200227 ± 0.0000012 | 5518.41097 ± 0.00020 |
| TrES-1 | 31 | 5.0 | 3.0300724 ± 0.0000004 | 3186.80703 ± 0.00012 |
| TrES-2 | 54 | 5.2 | 2.4706128 ± 0.0000003 | 3957.63574 ± 0.00011 |
| TrES-3 | 28 | 3.2 | 1.3061865 ± 0.0000002 | 4185.91110 ± 0.00008 |
| TrES-4 | 9 | 3.1 | 3.5539303 ± 0.0000019 | 4230.90575 ± 0.00043 |
| WASP-1 | 6 | 2.3 | 2.5199425 ± 0.0000014 | 3912.51531 ± 0.00032 |
| WASP-2 | 7 | 3.2 | 2.1522213 ± 0.0000004 | 3991.51536 ± 0.00018 |
| WASP-3 | 33 | 4.3 | 1.8468332 ± 0.0000004 | 4143.85194 ± 0.00017 |
| WASP-6 | 3 | 3.4 | 3.3609998 ± 0.0000011 | 4596.43342 ± 0.00013 |
| WASP-10 | 22 | 3.0 | 3.0927297 ± 0.0000003 | 4664.03803 ± 0.00006 |
| WASP-11/HAT-P-10 | 8 | 6.0 | 3.7224793 ± 0.0000007 | 4759.68753 ± 0.00011 |
| WASP-12 | 6 | 2.7 | 1.0914224 ± 0.0000003 | 4508.97683 ± 0.00019 |
| WASP-24 | 10 | 2.1 | 2.3412162 ± 0.0000014 | 5081.38033 ± 0.00010 |
| WASP-32 | 2 | 1.9 | 2.7186591 ± 0.0000024 | 5151.05460 ± 0.00050 |
| WASP-33 | 5 | 4.6 | 1.2198721 ± 0.0000003 | 4163.22465 ± 0.00022 |
| WASP-48 | 2 | 0.9 | 2.1436283 ± 0.0000041 | 5364.55120 ± 0.00027 |
| WASP-50 | 3 | 0.8 | 1.9550905 ± 0.0000022 | 5558.61277 ± 0.00020 |
| XO-1 | 25 | 5.0 | 3.9415052 ± 0.0000008 | 3808.91777 ± 0.00011 |
| XO-5 | 22 | 2.8 | 4.1877545 ± 0.0000016 | 4485.66876 ± 0.00028 |

TABLE 3

MODELED SYSTEM PARAMETERS

| Name | Limb Darkening Coeff. a | Limb Darkening Coeff. b | i [°] | a / R* | $R_p$ / R* |
|---|---|---|---|---|---|
| HAT-P-3 | 0.095 | 0.398 | $85.72^{+0.58}_{-0.51}$ | $9.21^{+0.56}_{-0.48}$ | $0.1093^{+0.0020}_{-0.0019}$ |
| HAT-P-4 | 0.102 | 0.366 | $86.01^{+2.74}_{-3.24}$ | $5.64^{+0.42}_{-0.74}$ | $0.0804^{+0.0049}_{-0.0051}$ |
| HAT-P-6 | 0.047 | 0.360 | $83.89^{+0.84}_{-0.74}$ | $6.77^{+0.51}_{-0.41}$ | $0.0970^{+0.0023}_{-0.0023}$ |
| HAT-P-12 | 0.221 | 0.313 | $88.46^{+0.99}_{-0.93}$ | $11.22^{+0.45}_{-0.69}$ | $0.1404^{+0.0026}_{-0.0026}$ |
| HAT-P-27/WASP-40 | 0.170 | 0.341 | $84.23^{+0.88}_{-0.88}$ | $9.11^{+1.01}_{-0.71}$ | $0.1344^{+0.0389}_{-0.0174}$ |
| HAT-P-32 (1) | 0.064 | 0.366 | $88.16^{+1.17}_{-1.03}$ | $5.98^{+0.10}_{-0.15}$ | $0.1531^{+0.0012}_{-0.0012}$ |
| HD 189733 | 0.187 | 0.332 | $86.05^{+0.43}_{-0.37}$ | $9.12^{+0.38}_{-0.32}$ | $0.1536^{+0.0025}_{-0.0026}$ |
| Qatar-1 | 0.195 | 0.340 | $84.81^{+0.82}_{-0.74}$ | $6.56^{+0.36}_{-0.32}$ | $0.1499^{+0.0023}_{-0.0030}$ |
| TrES-2 | 0.087 | 0.362 | $84.08^{+0.61}_{-0.67}$ | $7.93^{+0.57}_{-0.54}$ | $0.1295^{+0.0048}_{-0.0039}$ |
| TrES-4 | 0.065 | 0.372 | $82.36^{+1.77}_{-1.16}$ | $5.79^{+0.84}_{-0.47}$ | $0.0942^{+0.0030}_{-0.0035}$ |
| WASP-1 | 0.076 | 0.374 | $85.90^{+2.62}_{-2.37}$ | $5.24^{+0.32}_{-0.46}$ | $0.1069^{+0.0034}_{-0.0033}$ |
| WASP-2 | 0.170 | 0.341 | $85.18^{+1.26}_{-1.03}$ | $8.22^{+1.08}_{-0.82}$ | $0.1135^{+0.0053}_{-0.0056}$ |
| WASP-3 | 0.055 | 0.366 | $86.33^{+2.51}_{-2.73}$ | $5.48^{+0.33}_{-0.55}$ | $0.0951^{+0.0043}_{-0.0045}$ |
| WASP-6 | 0.119 | 0.359 | $88.60^{+0.91}_{-0.88}$ | $10.44^{+0.27}_{-0.46}$ | $0.1395^{+0.0014}_{-0.0013}$ |
| WASP-10 | 0.231 | 0.311 | $88.64^{+0.91}_{-0.91}$ | $11.81^{+0.44}_{-0.72}$ | $0.1598^{+0.0040}_{-0.0038}$ |
| WASP-11/HAT-P-10 (2) | 0.222 | 0.320 | $89.24^{+0.52}_{-0.69}$ | $12.11^{+0.18}_{-0.38}$ | $0.1255^{+0.0020}_{-0.0018}$ |
| WASP-11/HAT-P-10 (3) | 0.222 | 0.320 | $89.08^{+0.64}_{-0.86}$ | $12.28^{+0.28}_{-0.60}$ | $0.1256^{+0.0027}_{-0.0028}$ |
| WASP-32 | 0.071 | 0.362 | $85.00^{+1.58}_{-1.29}$ | $7.71^{+1.08}_{-0.82}$ | $0.1030^{+0.0031}_{-0.0033}$ |
| WASP-33 (4) | 0.016 | 0.359 | $83.24^{+3.71}_{-2.14}$ | $3.31^{+0.20}_{-0.16}$ | $0.1022^{+0.0027}_{-0.0028}$ |
| WASP-24 | 0.076 | 0.365 | $86.42^{+2.37}_{-2.83}$ | $7.08^{+0.66}_{-1.12}$ | $0.0998^{+0.0039}_{-0.0040}$ |
| WASP-48 | 0.252 | 0.301 | $85.06^{+3.05}_{-3.13}$ | $5.44^{+0.57}_{-0.76}$ | $0.0988^{+0.0051}_{-0.0049}$ |
| WASP-50 | 0.128 | 0.358 | $85.31^{+0.82}_{-0.68}$ | $7.90^{+0.56}_{-0.46}$ | $0.1347^{+0.0037}_{-0.0037}$ |
| XO-5 | 0.137 | 0.356 | $87.24^{+1.41}_{-1.14}$ | $10.12^{+1.01}_{-1.01}$ | $0.1022^{+0.0033}_{-0.0031}$ |

[1] Parameters for the 2011 October 09 lightcurve.
[2] Parameters for the 2009 November 26 J-band lightcurve which exhibits starspot activity.
[3] Parameters for the 2011 October 08 J-band lightcurve.
[4] Exhibits stellar variability and uneven baseline outside the transit.

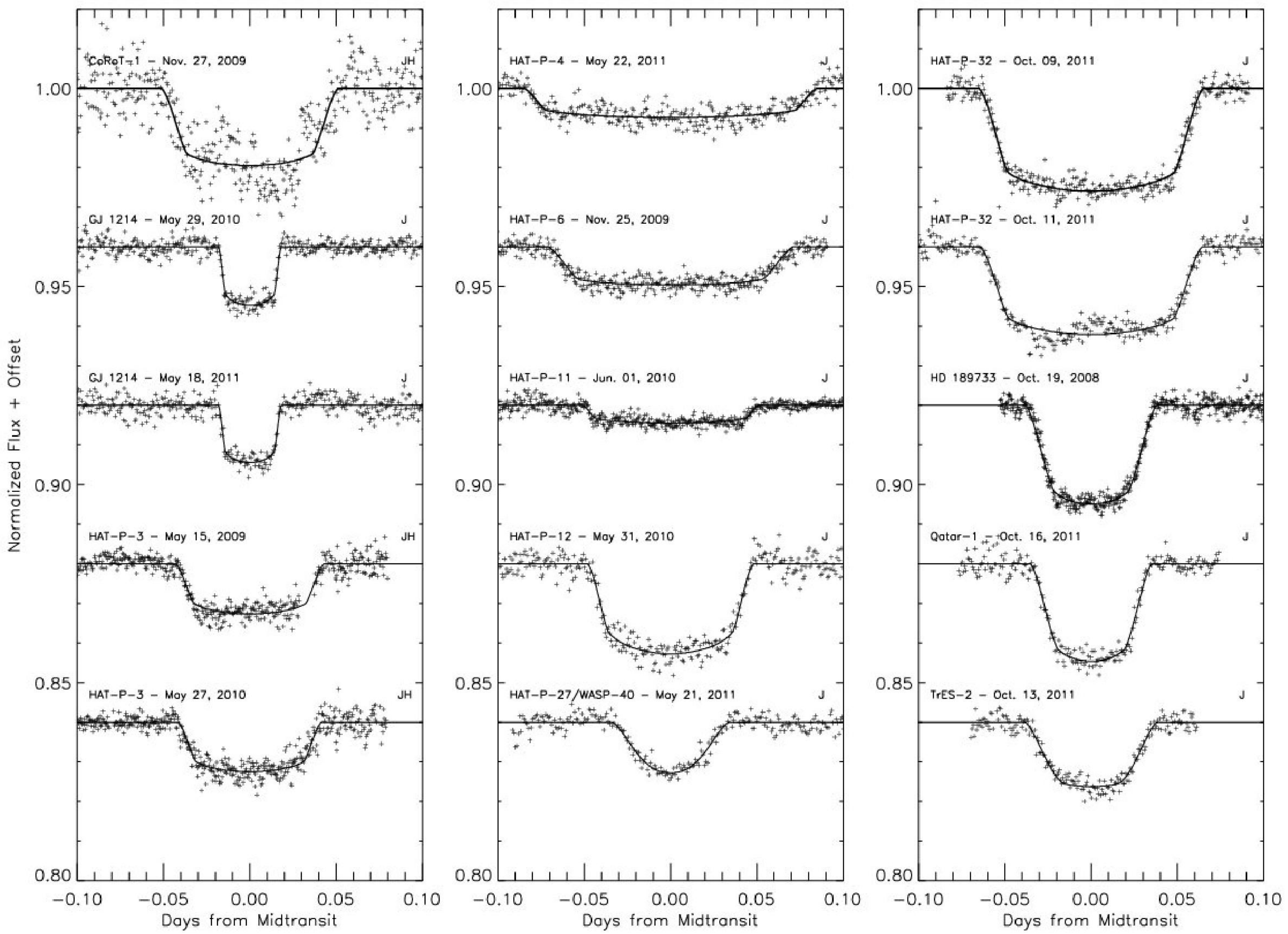

Figure 1a: Exoplanet Transits observed in the near-IR with the KPNO 2.1m telescope

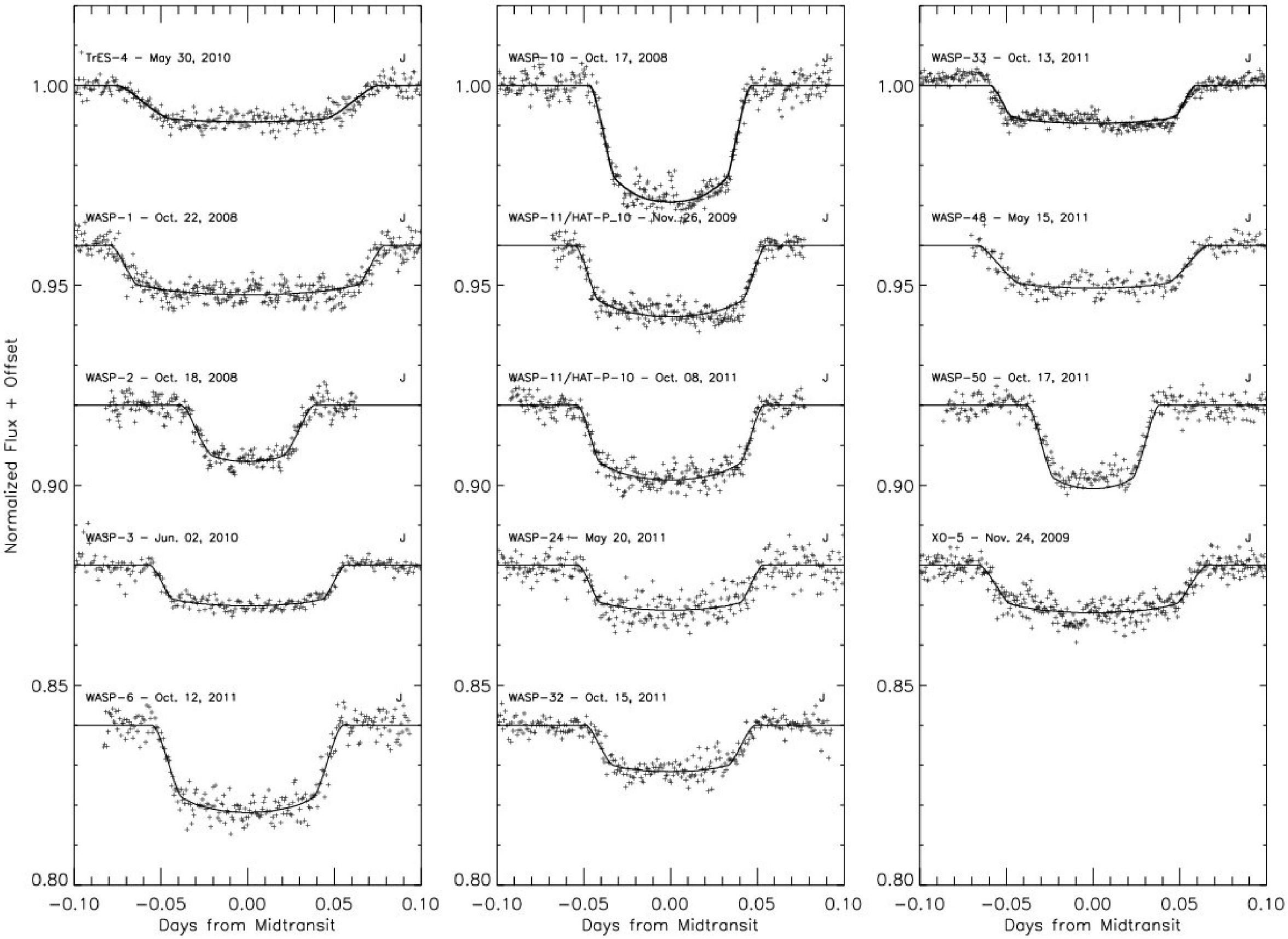

Figure 1b: Exoplanet Transits observed in the near-IR with the KPNO 2.1m telescope

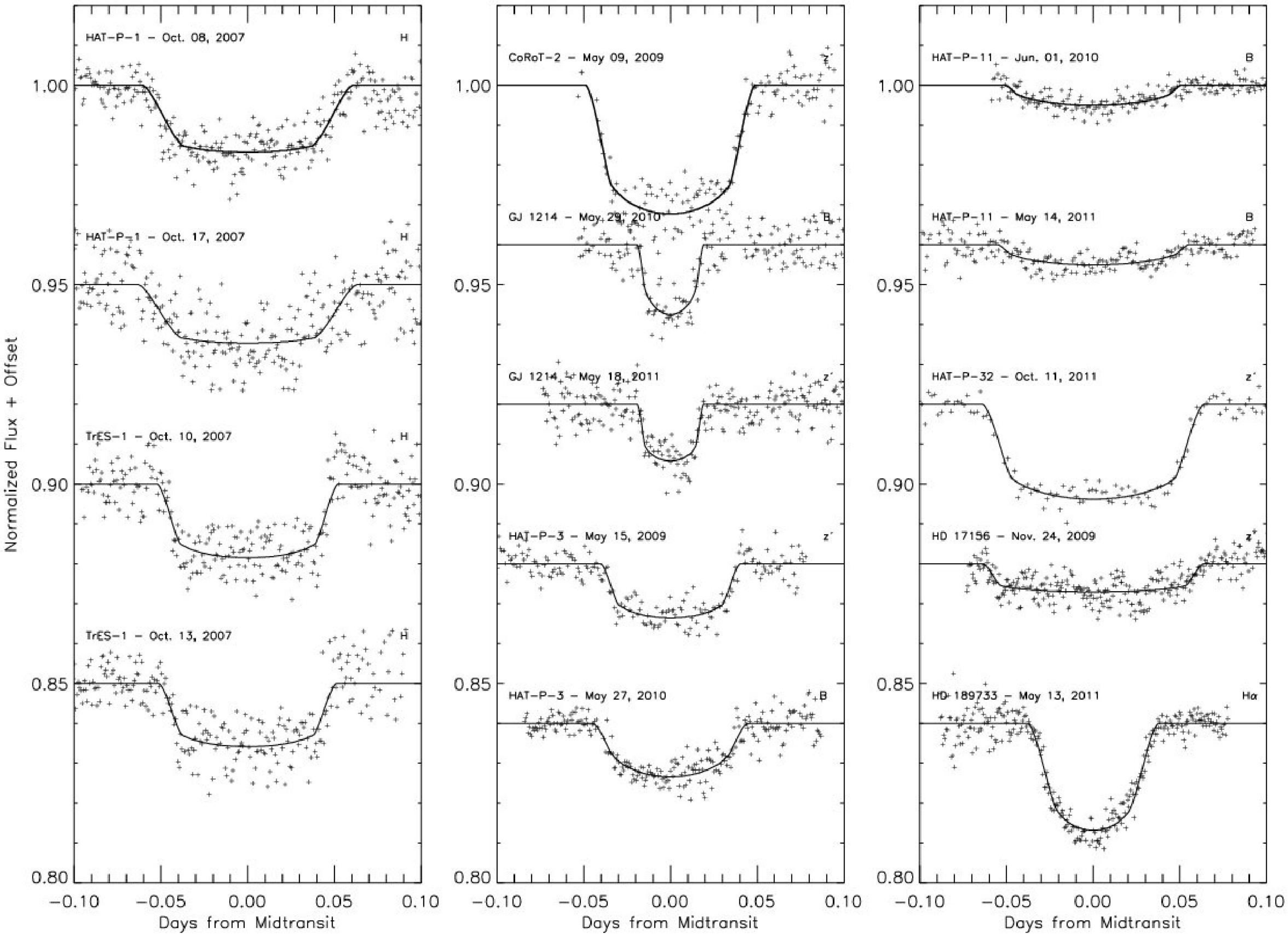

Figure 2a: Exoplanet Transits observed in the near-IR with the KPNO National Solar Observatory McMath-Pierce 2.0m telescope (left column) and in the visible with the KPNO Visitor Center 0.5m telescope.

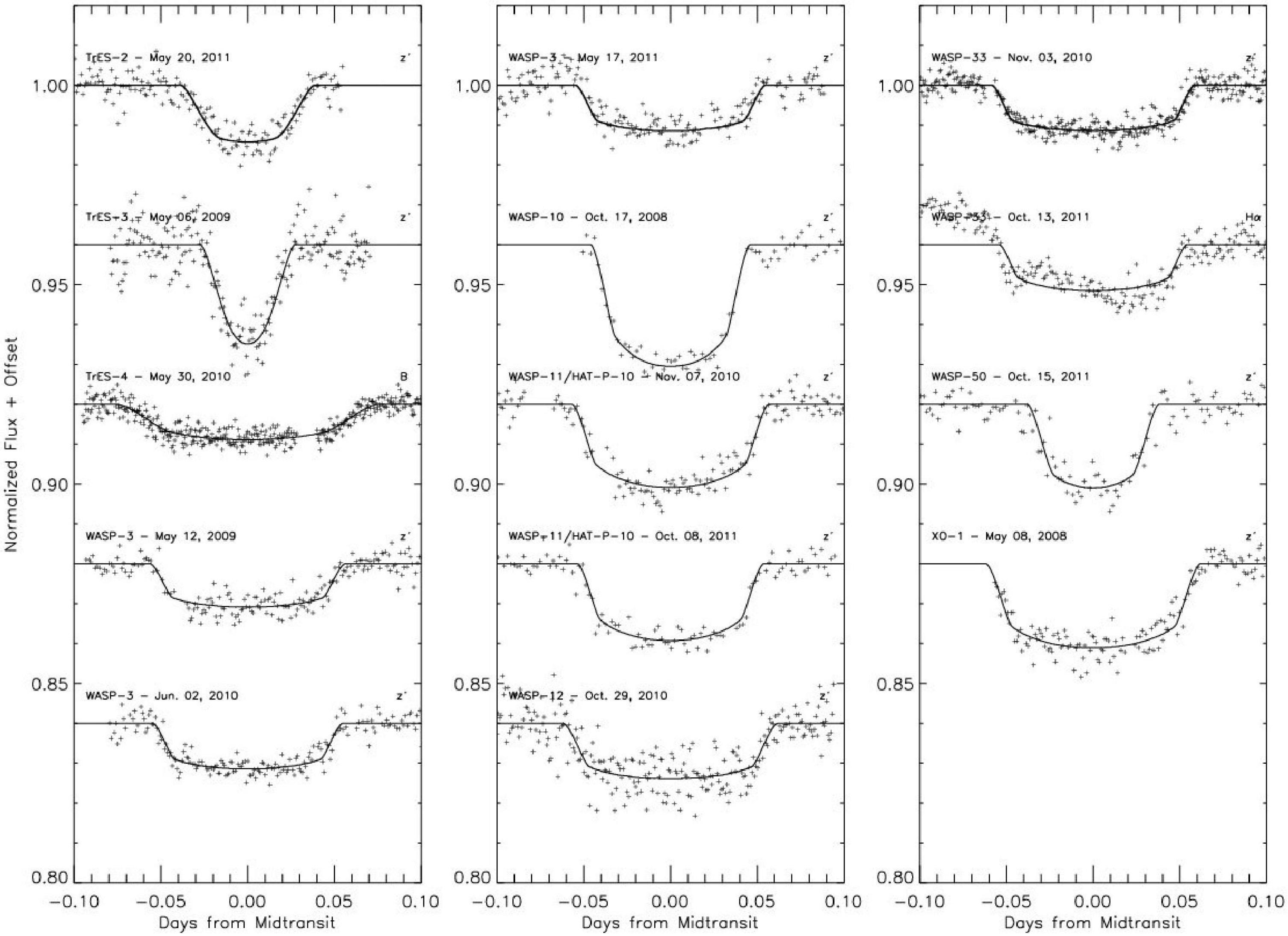

Figure 2b: Exoplanet Transits observed in the visible with the KPNO Visitor Center 0.5m telescope.

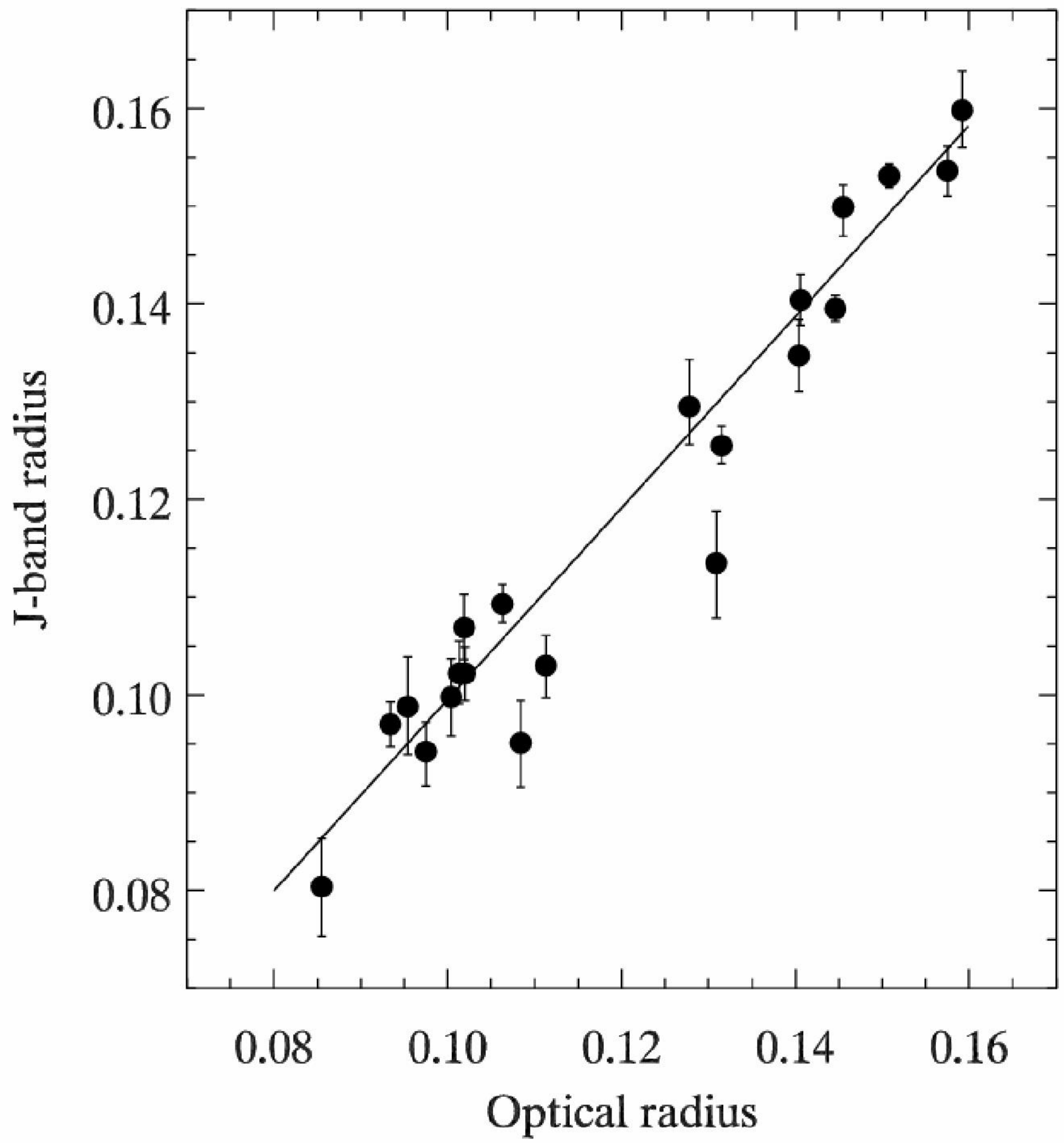

Figure 3: Ratio of planetary to stellar radius, $R_p/R_*$, from MCMC fitting to our J-band transit curves (ordinate, from Table 3), compared to published optical radii for the same planets. We exclude HAT-P-27/WASP-40 due to its much larger error. The solid line is a least squares fit (see text), and it differs negligibly from the null hypothesis that the derived planetary radii are independent of wavelength (see text).

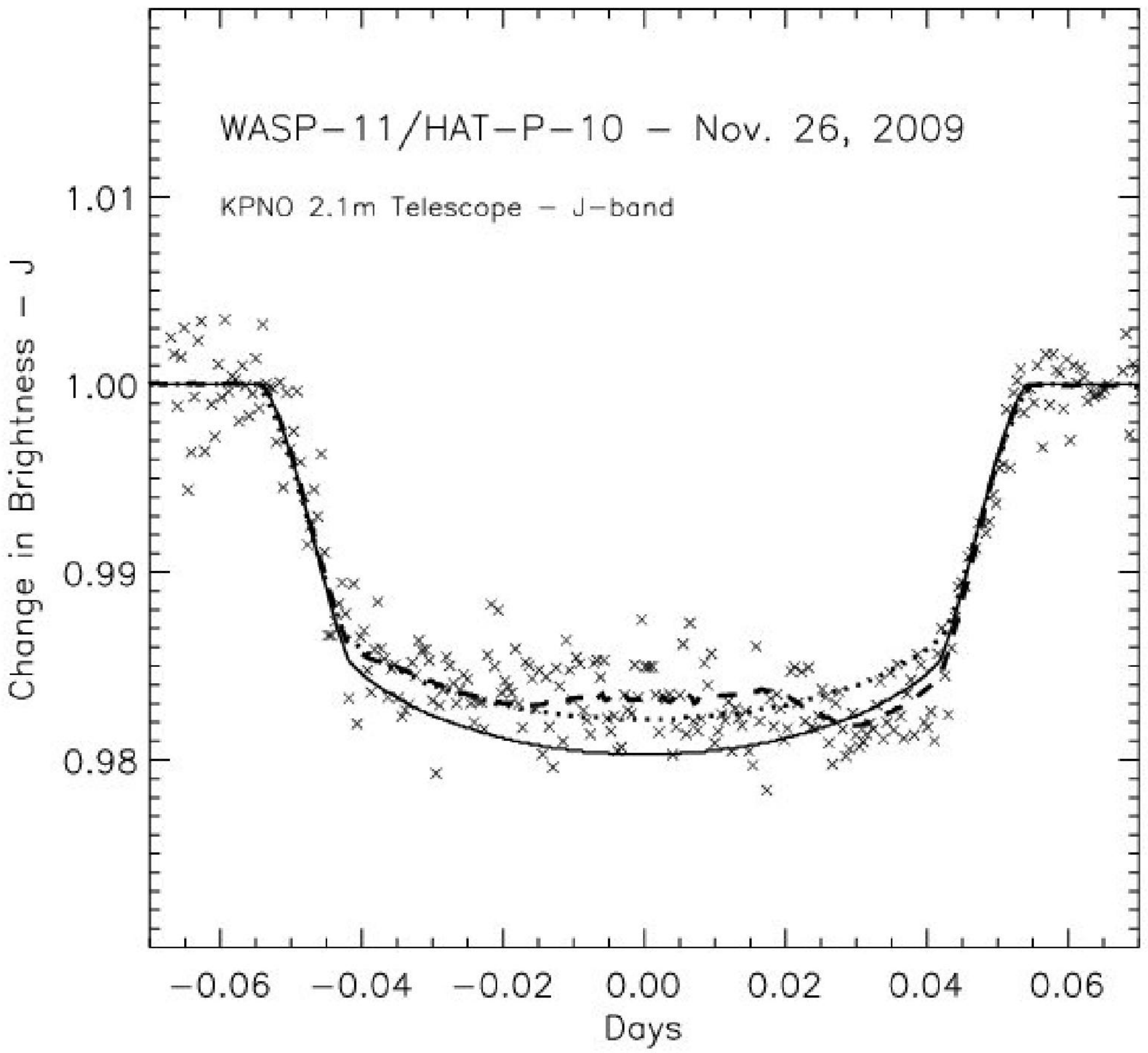

Figure 4: Observed transit of HAT-P-10/WASP-11 on 2009 November 26 at the KPNO 2.1m telescope through a J-Band filter. The solid line is the model derived from the literature parameters. The dotted line is our fit to the entire depth of the transit. The dashed line is the literature model but with part of the stellar surface covered by starspots with temperature ~6% lower than the effective temperature of the star (4920 K).